# Quantitative T2 Estimation Using Radial Turbo Spin Echo Imaging

Mahesh B Keerthivasan[1,2], Ali Bilgin[1,3], and Maria I Altbach[2]

Electrical and Computer Engineering, University of Arizona, Tucson, USA
Medical Imaging, University of Arizona, Tucson, USA
Biomedical Engineering, University of Arizona, Tucson, USA

## INTRODUCTION

T2-weighted (T2w) MRI is key in abdominal imaging and used in the characterization of a variety of pathologies. T2-weighted imaging is particularly important in the characterization of hemangiomas and cysts, the most common benign liver lesions; these benign lesions are hyperintense compared to liver with distinct higher signal intensity in T2w images compared to metastases.

Abdominal T2-weighted imaging can be performed with multi-shot or single-shot turbo spin-echo (TSE) pulse sequences. The multi-shot TSE pulse sequence offers higher spatial resolution but image quality with conventional Cartesian acquisitions is deteriorated by artifacts due to flow and motion, undermining the intrinsic spatial resolution and impairing detection of small tumors. Single shot TSE methods are robust to motion but the signal loss and blurriness associated with T2 filtering due to long echo trains affects spatial resolution. Single shot acquisitions can be performed with breath holding or free breathing since the acquisition of each slice is < 1 sec thus, most physiological motion is minimized. Multi shot TSE acquisitions can be performed during breath holding or free breathing using respiratory triggering methods; the latter has the advantage of improved signal-to-noise ratio (SNR) but comes at the expense of longer acquisition times, in particular when the respiration trace is not consistent.

**RADIAL TURBO SPIN ECHO IMAGING: PULSE SEQUENCE**

The Radial Turbo Spin Echo (RADTSE) pulse sequence has been proposed for T2-weighted imaging and T2 mapping [1-5] as an alternative to conventional Cartesian TSE methods. Figure 1 shows a RADTSE pulse sequence. Compared to a Cartesian trajectory where motion is observed as a directional ghosting, a radial trajectory distributes the effects of motion along two dimensions [6,7]. Hence, the pulse sequence allows the generation of images with high spatial and temporal resolution while being motion robust.

At each echo time, the $G_x$ and $G_y$ gradients together acquire one radial spoke (view) in k-space:

$$G_x = G \cos \phi$$

$$G_y = G \sin \phi$$

where $\phi$ is the azhimuth angle.

Due to the radial trajectory, every radial view samples the center of k-space. The acquired radial k-space data contains views from all the TE times. Reconstructing this dataset results in a composite image with an average T2w contrast and can be used as an anatomical reference. The acquired k-space data can also be partitioned into undersampled k-space data corresponding to each TE which can be reconstructed to generate TE images. Thus, RADTSE allows the generation of a series of co-registered images at different T2-weighted contrasts from a single data set.

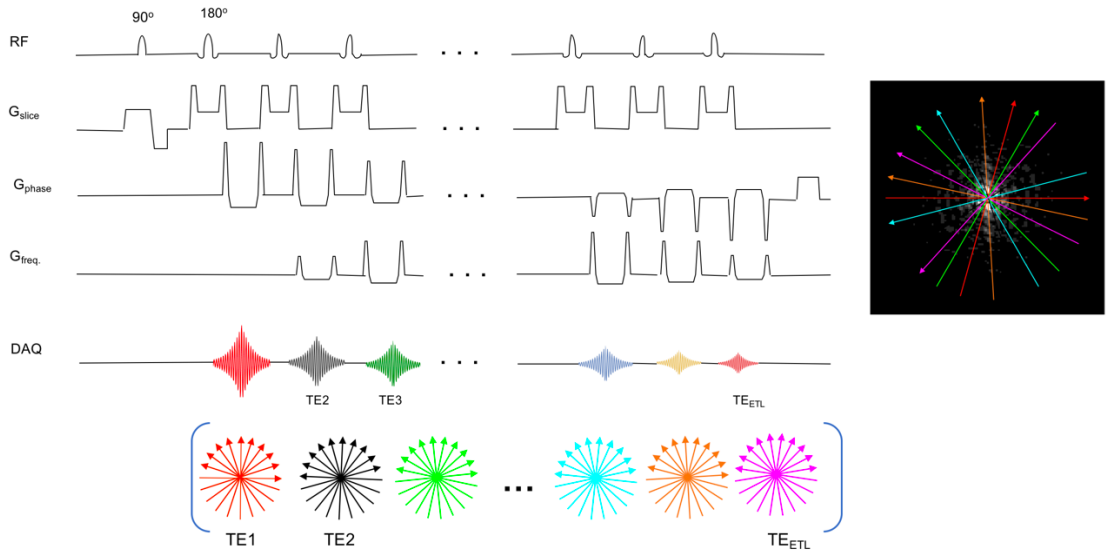

**Figure 1:** Radial Turbo Spin Echo pulse sequence. The sequence consists of turbo spin echo readouts that sample k-space in a radial manner. During each echo time one radial spoke is acquired and k-space is filled in consecutive TRs. Since the center of k-space is sampled at each echo time, the acquired data can be partitioned into under-sampled radial data corresponding to each echo time.

**Radial View Ordering**

In a TSE pulse sequence, the echo train length and the number of shots are known a priori for each kz partition allowing the design of efficient k-space coverage schemes. An efficient view ordering scheme (i) ensures uniform coverage of k-space for each TE and (ii) distributes the views to reduce artifacts from T2 decay. A sequential view ordering scheme could be used to acquire k-space data where the angular ordering of views is given by $\phi_{j,k} = (k-1) * ETL * \frac{\pi}{N} + j\frac{\pi}{N}$, where $j$ is the echo index and $k$ is the TR or shot index and $N$ is the total number of radial views to be acquired. While this view ordering would provide uniform k-space coverage, the T2 signal decay during each echo train causes a low frequency modulation along the acquired angles resulting in streak artifacts. Moreover, the grouping of angles in each TR will exacerbate errors arising from motion during a TR.

As an alternative, a bit reversed view ordering scheme [9] where the echoes within a small angular section are ordered to yield a rapidly varying T2 decay pattern (to reduce streaks) while providing uniform k-space coverage for each TE was proposed for 2D radial TSE. angular ordering of radial views for the bit-rev method is determined using:

$$\phi_{j,k} = \left\{ \left[ (m_j - 1) + (j-1)\frac{N}{ETL} + (k-1)ETL \right] \Delta\theta \right\} \bmod \pi$$

where *j* is the index for the echo number within an echo train, $m_j$ is the bit-rev index and depends on the ETL; for ETL=16, $m_j$ = [1,9,5,13,3,11, …., 16] and for ETL=32, $m_j$ = [1, ….32]. *N* is the number of radial views, *k* is the shot (or TR period) index, and $\Delta\theta = \frac{\pi}{N}$ is the angular increment between two adjacent views. The bit-reverse view ordering provides an ideal point spread function (PSF) when taking into account the change in contrast due to T2 decay reducing image artifacts when data from all TEs are used to reconstruct an image. With the bit-reverse order the radial views for each TE are evenly distributed over 2pi, an important aspect for T2 mapping. However, the bit-reversed scheme restricts the ETL to powers of 2 thereby reducing flexibility on the choice of ETL.

Another alternative is the golden angle view ordering scheme [9], widely used in gradient echo imaging. The golden angle view ordering does not restrict the choice of ETL, however, at very high acceleration rates (such as 3-4 views per TE) it yields a non-uniform coverage of k-space.

Recently, a more flexible view ordering was proposed for 3D radial TSE [10] that combines flexibility in the choice of ETL and uniform sampling of k-space for each TE. The angle acquired at $TE_j$ and the $TR_k$ is given by:

$$\phi_{j,k} = \left(mod(k+j,\rho) * \frac{\pi}{\rho}\right) + (ETL - j) * \left(\frac{\pi}{N}\right); j = 1,3, \ldots, ETL - 1$$

$$\phi_{j,k} = \left(mod(k+j,\rho) * \frac{\pi}{\rho}\right) + j * \left(\frac{\pi}{N}\right); j = 0,2, \ldots, ETL - 1,$$

where, $N$ is the number of radial views and $\rho = N/ETL$.

Figure 2 shows the radial coverage and point spread function for the bit-reversed, golden angle, and pseudo-random view ordering schemes. The bit-reversed and pseudo-random view orderings have uniform radial coverage with minimum side lobe amplitude.

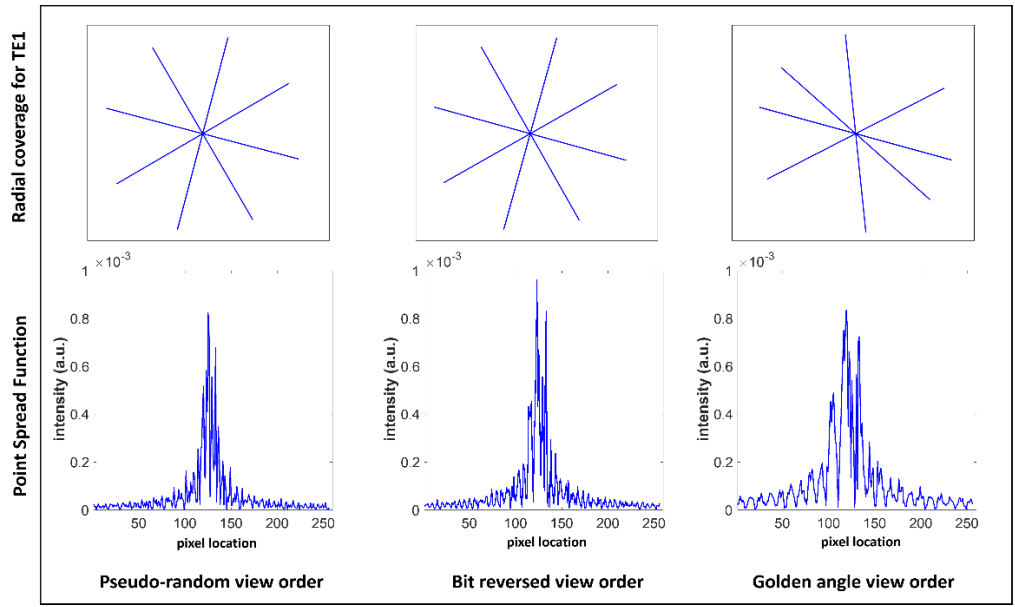

**Figure 2:** Radial coverage and point spread function for the bit-reversed, golden angle, and pseudo-random view ordering schemes.

## RADIAL TURBO SPIN ECHO IMAGING: IMAGE RECONSTRUCTION

The highly undersampled k-space data corresponding to the individual TEs can be reconstructed using either an echo sharing approach [2] or using iterative reconstruction algorithms [3,11,12] (Figure 3).

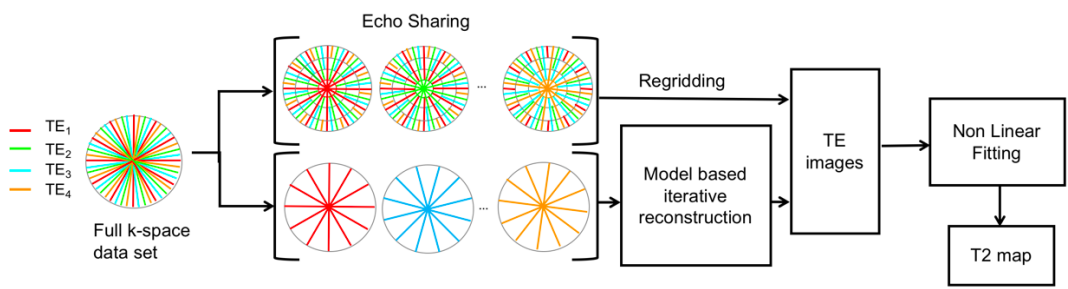

**Figure 3:** Undersampled k-space data at each TE can be reconstructed using either an echo sharing approach or model based iterative algorithms to generate a series of echo images. The T2 map is generated by fitting the echo images to the underlying signal model.

### Echo Sharing

The echo sharing algorithm for RADTSE data [2] is based on the property that the image contrast is determined by the low frequency region in k-space. Since the center of k-space is well sampled for each TE in a radial acquisition, this technique retains data corresponding to a specific TE at the center of k-space up to a radius determined by the Nyquist sampling criteria (Nyquist radius). Beyond the Nyquist radius, the radial views acquired at adjacent TE times are included in a multi-tier fashion. This is illustrated in Figure 4 for a data with ETL=8. In order to generate images with T2w contrast corresponding to the $j^{th}$ TE time, the innermost tier contains radial views from only $TE_j$. The subsequent tiers are filled by borrowing data from other TEs. Since the effective TE is determined by the TE of k-space data in the innermost tier, the T2w contrast is retained at each echo time. However, the mixing of data at high spatial frequencies introduces blurring of small structures and fine edges. These TE images can then be used for the generation of high resolution T2 maps using a non-linear fitting, thus providing quantitative T2 information along with qualitative T2w images from a single acquisition.

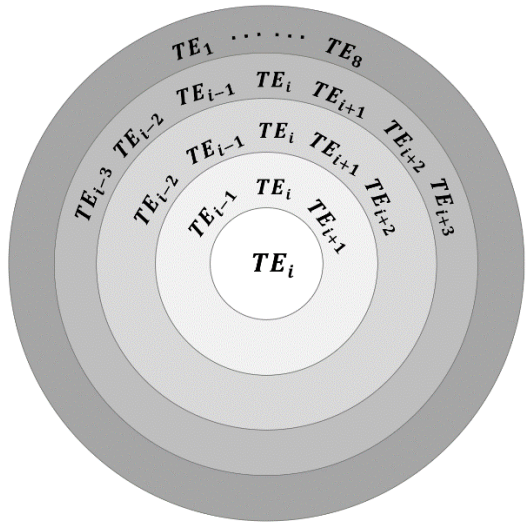

**Figure 4:** An illustration of the echo sharing algorithm for an echo train length of 8 is shown here. For a T2w contrast corresponding to the $j^{th}$ TE, the innermost tier contains radial views only from TE$_j$. The second tier consists of radial views from neighboring TEs. This iterative mixing of radial views is continued till the last tier, which contains radial views from all 8 TEs.

**Subspace Constrained Reconstruction**

In order to overcome the effects of mixing of TE data from echo sharing, iterative reconstruction algorithms have been proposed [11,13] for the direct reconstruction of T2 maps. These techniques estimate T2 together with the equilibrium magnetization ($I_0$) and the B1 field map for all voxels by solving the following model-based reconstruction problem:

$$\widehat{T2} = arg\ min_{T2,I_0,B1} \sum_{j=1}^{ETL} \left\Vert FS\{f_j(I_0, T2, B1)\} - K_j \right\Vert_2^2$$

where $S$ is the coil sensitivity matrix and $F$ is the non-uniform fast Fourier transform [14] operator corresponding to the acquired trajectory for the $j^{th}$. $K_j$ is the undersampled k-space data at the $j^{th}$ TE, and $f_j$ is the signal at the $j^{th}$ TE obtained using a TSE signal model and is a function of $I_0, T2\ and\ B1$. Due to the non-linearity of the function $f(.)$, solutions to the optimization problem are sensitive to scale mismatch between $I_0$ and T2 [3] and choice of the scaling factor affects accuracy of the estimated T2.

Alternately, subspace constrained reconstruction algorithms [3,12,15] have been proposed for the reconstruction of individual TE images. These echo images can subsequently be used to estimate the T2 values. Subspace based techniques exploit the signal redundancy across the TE images such that the information in $ETL$ number of images can be well approximated in a lower dimensional subspace. Let $X$ be a $ETL \times N$ matrix such that the rows are the different echoes and the columns are the $N$ pixels in each TE image. $X$ can then be approximated as $X \approx \Phi M$, where $\Phi$ is a $L$ dimensional principal component (PC) subspace basis such that $L << ETL$ and $M$ corresponds to the principal component coefficients. The subspace basis $\Phi$ is generated by simulating the TSE signal using the SEPG model for a range of T2 and B1 values. Figure 5(A) shows an ensemble of T2 signal evolutions simulated using the SEPG model for ETL = 16 along with the first four principal components obtained using PC analysis (Figure 5(B)). Note that these four PCs are sufficient to represent the signal evolution with high accuracy. The effect of subspace projection is illustrated in Figure 6 using data synthesized from a digital brain phantom. Five representative TE images (out of 16) are shown along with the first four PC coefficient images. Note that the PC coefficient images are spatially correlated allowing to be well represented in a transform domain.

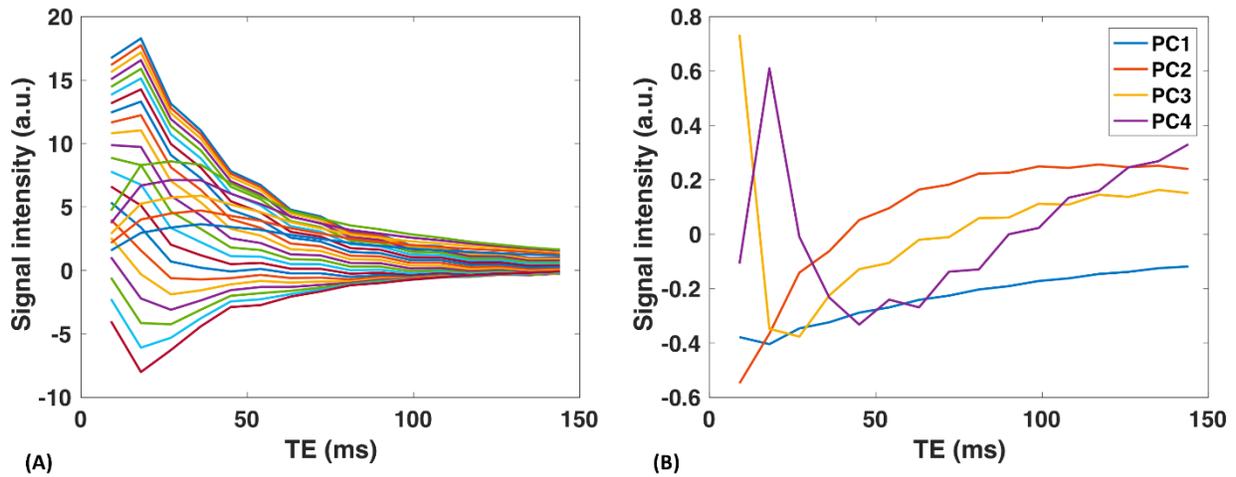

**Figure 5:** An ensemble of T2 signal evolution curves (A) simulated using the SEPG model for an echo train length of 16 along with the first four principal components (B) obtained using the principal component analysis are shown here. The signal curves are generated for a range of T2 and B1 values.

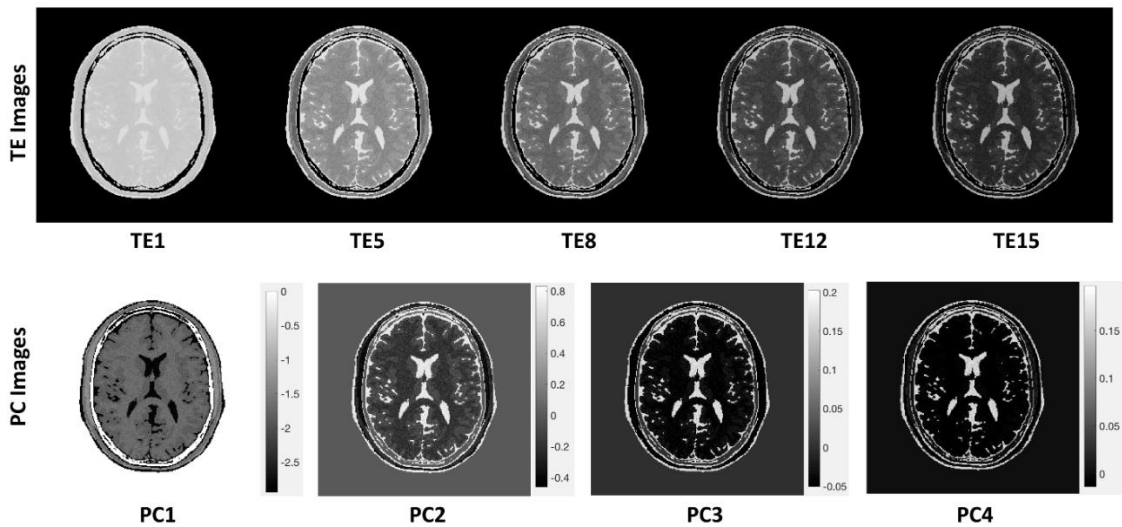

**Figure 6:** An axial image of a digital brain phantom along with representative T2w images at five different TEs are shown in the top row. The bottom row shows the projection of these images on the first four principal components. Note that the PC coefficient images are spatially correlated, allowing to be well represented in a transform domain.

We define the k-space encoding operator as $E = FS$. Here, $S$ is the coil sensitivity matrix that modulates the TE images and $F$ is the non-uniform fast Fourier transform operator corresponding to the acquired trajectory for the different echo times. The forward signal model used to generate k-space can be written as $K = EX \approx E\Phi M$, where $K$ is the multi-coil k-space data corresponding to the temporal points. The reconstruction attempts to jointly recover the PC coefficients by solving the following optimization problem:

$$\widehat{M} = arg\,min_M \left\|E\Phi M - K\right\|_2^2 + \lambda R(M)$$

The second term in the above equation is a penalty term used to exploit the spatial compressibility of the PC coefficient images weighted by a regularization parameter $\lambda$. The penalty function $R(.)$ could be a a finite difference operator [3,12,16] or a low rank operator [16-18] among other possibilities.

**Quantitative T2 Estimation**

The reconstructed TE images are perfectly co-registered with each other and can be used to generate a quantitative T2 image or map. This is achieved by fitting the temporal signal for each pixel to the TSE signal model as shown below:

$$\widehat{T2}, \widehat{B1}, \widehat{I_0} = arg\,min_{T2,I_0,B1} \sum_{j=1}^{ETL} \left\|f_j(I_0, T2, B1, \theta_0, \ldots, \theta_j, j) - s_j\right\|_2^2$$

where, $s_j$ is the signal intensity at the $j^{th}$ echo and $f(.)$ is the underlying signal model and can be either the single exponential function or the EPG / SEPG model [19].

While the above optimization problem can be solved using an iterative non-linear least squares algorithm, a more computationally efficient approach is to use a pattern matching technique [4,11,20]. This technique matches the temporal signal from the images to a pre-determined dictionary of temporal signals obtained by simulating the forward model.

**EXPERIMENTS AND RESULTS**

**Phantom Imaging**

In order to evaluate efficiency of the RADTSE technique for quantitative T2 estimation, it was evaluated using a set of agarose gel phantoms with different T2 times. Data were acquired using RADTSE on a 3T Siemens scanner (Skyra, Siemens Healthcare, Erlangen, Germany) with ETL = 32, echo spacing = 7.2ms and 192 radial views. Reference data were acquired using a Cartesian Spin Echo pulse sequence as this method is not affected by indirect echoes. SE data were acquired with matrix size = 256x256, TR = 1500ms, and it was repeated for different TE values in increments of 7ms.

The mean T2 values and standard deviation for the four phantoms estimated by the RADTSE pulse sequence with both the mono exponential fit and the SEPG dictionary fit are compared to the spin echo reference in Table 1. Note that the larger bias in the estimates from the mono-exponential fit shows the need to account for the stimulated echoes in the fit.

Table 1: Mean T2 Estimates (ms)

| Spin-Echo Reference | Mono Exponential Fit | SEPG Fit |
|---|---|---|
| 76.94 ± 2.54 | 101.68 ± 2.92 | 79.2 ± 2.20 |
| 88.15 ± 2.15 | 117.17 ± 2.12 | 92.4 ± 1.61 |
| 133.23 ± 4.25 | 145.36 ± 3.65 | 136.36 ± 2.43 |
| 158.23 ± 3.12 | 202.43 ± 3.21 | 160.29 ± 2.66 |

**In vivo Imaging**

Radial FSE data was acquired on normal volunteers at 3T Siemens scanner. The imaging parameters were ETL = 32, echo spacing = 6.2ms and 192 radial views. The range of T2 values used for principal component dictionary was 20 – 1000ms.

Figure 7 and Figure 8 show axial slices of the abdomen acquired on healthy volunteers using the RADTSE pulse sequence. The images correspond to different echo times and show different T2 weighting. The composite image generated from all the acquired radial views exhibits an average T2w contrast. Also shown is the quantitative T2 map obtained by fitting the TE images.

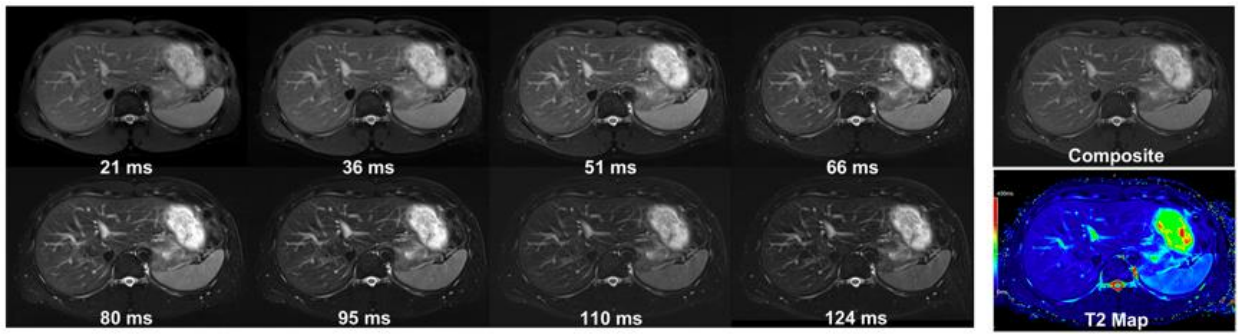

**Figure 7:** Representative T2w images (8 out of 21 TEs) and a T2 map of the abdomen generated using RADTSE for a normal volunteer. The composite image with an average T2w contrast is also shown.

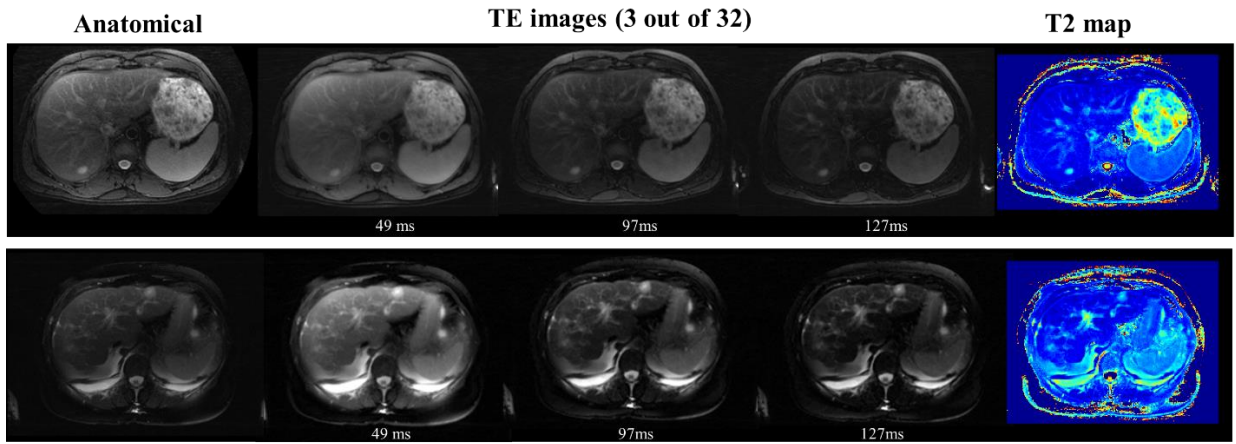

**Figure 8:** Representative T2w images (3 out of 32 TEs) and T2 maps of the abdomen from two normal volunteers along with the composite anatomical reference image.

**CONCLUSION**

The use of a Radial Turbo Spin Echo pulse sequence with model based T2 estimation can allow efficient quantitative T2 estimation. This technique allows generation of co-registered T2-weighted images at multiple TE times along with the T2 map. The use of T2 fitting models that account for RF slice profile variations improves the estimation accuracy. In vivo experiments indicate the utility of this technique for abdominal T2 mapping with the ability to cover the whole abdomen in up to 3 breath-holds.

**REFERENCES**

[1] Altbach MI, Outwater EK, Trouard TP, Krupinski EA, Theilmann RJ, Stopeck AT, Kono M, and Gmitro AF. Radial fast spin-echo method for T2-weighted imaging and T2 mapping of the liver. Journal of Magnetic Resonance Imaging, 16(2):179–189, 2002.

[2] Altbach MI, Bilgin A, Li Z, Clarkson EW, Trouard TP, and Gmitro AF. Processing of radial fast spin-echo data for obtaining T2 estimates from a single k-space data